\documentclass[journal]{IEEEtran}
\ifCLASSINFOpdf

\else
 
\fi

\usepackage{cite}
\usepackage{amsmath,amssymb,amsfonts}
\usepackage{algorithmic}
\usepackage{graphicx}
\usepackage{textcomp}
\usepackage{cite}
\usepackage{multirow}
\usepackage{balance}
\usepackage{colortbl}
\usepackage{hyperref}
\usepackage{url}
\usepackage{caption}
\usepackage{comment}
\usepackage{subcaption}
\definecolor{Gray}{gray}{0.90}
\usepackage{hyperref}
\usepackage{setspace} 
\usepackage{soul}

\begin{document}

\title{Early Myocardial Infarction Detection over Multi-view Echocardiography}

\author{Aysen Degerli, 
        Serkan Kiranyaz, 
        Tahir Hamid, 
        Rashid Mazhar, 
        and Moncef Gabbouj

\thanks{This work was supported by the NSF-Business Finland Center for Visual and Decision Informatics (CVDI) Advanced Machine Learning for Industrial Applications (AMaLIA) project under Grant 4183/31/2021.}

\thanks{Aysen Degerli \textit{(aysen.degerli@tuni.fi)} and Moncef Gabbouj \textit{(moncef.gabbouj@tuni.fi)} are with the Faculty of Information Technology and Communication Sciences, Tampere University, Tampere, Finland. }
\thanks{Serkan Kiranyaz \textit{(mkiranyaz@qu.edu.qa)} is with the Department of Electrical Engineering, Qatar University, Doha, Qatar.}
\thanks{Tahir Hamid and Rashid Mazhar are with the Hamad Medical Corporation, Doha, Qatar.}
}

\markboth{Journal of \LaTeX}%
{Shell \MakeLowercase{\textit{et al.}}: A Sample Article Using IEEEtran.cls for IEEE Journals}

\maketitle

\begin{abstract}
Myocardial infarction (MI) is the leading cause of mortality in the world that occurs due to a blockage of the coronary arteries feeding the myocardium. An early diagnosis of MI and its localization can mitigate the extent of myocardial damage by facilitating early therapeutic interventions. Following the blockage of a coronary artery, the regional wall motion abnormality (RWMA) of the ischemic myocardial segments is the earliest change to set in. Echocardiography is the fundamental tool to assess any RWMA. Assessing the motion of the left ventricle (LV) wall only from a single echocardiography view may lead to missing the diagnosis of MI as the RWMA may not be visible on that specific view. Therefore, in this study, we propose to fuse apical 4-chamber (A4C) and apical 2-chamber (A2C) views in which a total of 12 myocardial segments can be analyzed for MI detection. The proposed method first estimates the motion of the LV wall by Active Polynomials (APs), which extract and track the endocardial boundary to compute myocardial segment displacements. The features are extracted from the A4C and A2C view displacements, which are concatenated and fed into the classifiers to detect MI. The main contributions of this study are 1) creation of a new benchmark dataset by including both A4C and A2C views in a total of 260 echocardiography recordings, which is publicly shared with the research community, 2) improving the performance of the prior work of threshold-based APs by a Machine Learning based approach, and 3) a pioneer MI detection approach via multi-view echocardiography by fusing the information of A4C and A2C views. Experimental results show that the proposed method achieves 90.91\% sensitivity and 86.36\% precision for MI detection over multi-view echocardiography. The software implementation is shared at \href{https://github.com/degerliaysen/MultiEchoAI}{https://github.com/degerliaysen/MultiEchoAI}.
\end{abstract}


\IEEEpeerreviewmaketitle

\section{Introduction}
\IEEEPARstart{M}{yocardial} infarction (MI) is caused by the death of myocardial cells subsequent to ischemia due to the blockage of coronary arteries. Presentation of MI is generally evident with shortness of breath, pain around the chest, shoulders, back, and arms \cite{universal}. However, these symptoms may not occur in the early stages of MI. Due to the blockage of the coronary artery and deprivation of blood supply, there is progressive damage to the affected part of the myocardium. Hence, it is critical to make an early detection of MI, to limit and prevent death \& disability. Currently, the diagnosis of MI is based upon a time-consuming method of serial observations of electrocardiography (ECG), blood level of cardiac enzymes, and imaging techniques \cite{thygesen2012third}. At the outset of MI, ECG is an insensitive tool with $0.77$ predictive value in ruling out MI \cite{grande1980optimal}. Moreover, human error leads to misdiagnosis of ischemic ECG changes in $12-16\%$ cases of MI \cite{goldman1988computer, qamar1999goldman, masoudi2006implications}. Furthermore, cardiac biomarkers take time to evolve to a diagnostic level. After the onset of MI, the high sensitivity cardiac troponin (hs-cTn) starts to rise in $3$ hours and it needs a repeat sample at least $6$ hours after onset of chest pain to quantify according to the American Heart Association definition of MI \cite{macrae2006assessing}. Therefore, the most convenient tool to diagnose and assess MI in its early stages is echocardiography, which has easy accessibility, low cost, and lowest risk compared to other cardiac imaging options \cite{chatzizisis2013echocardiographic, gottdiener2004american}.

Two-dimensional (2D) echocardiography was first introduced in the late $1950$s, which is a non-invasive ultrasound imaging technique that monitors the heart in real-time \cite{EDLER20041565, gottdiener2004american}. The early detection of MI can be performed by evaluating the regional wall motion abnormality (RWMA) in 2D echocardiography, where the abnormalities caused by MI can be detected as a region of weaker motion of the myocardium. However, the assessment of RWMA is highly subjective and variant among experts \cite{porter2018clinical}. Moreover, the echocardiography recordings are generally subject to a high level of noise with low image quality, where the left ventricle (LV) wall is mostly unrecognizable. Thus, visual assessment of the RWMA highly depends on the expertise of the echocardiographist and the quality of the echocardiography recordings. Therefore, in order to achieve reliable MI detection, computer-aided diagnosis techniques are developed to help cardiologists in the diagnosis. Consequently, motion estimation algorithms are utilized to assess and quantify the RWMA in echocardiography. Several approaches that are popular for estimating myocardial motion are optical flow methods, deformation imaging, and active contours. 

\begin{figure}[t!]
    \centering
    \includegraphics[width=\linewidth]{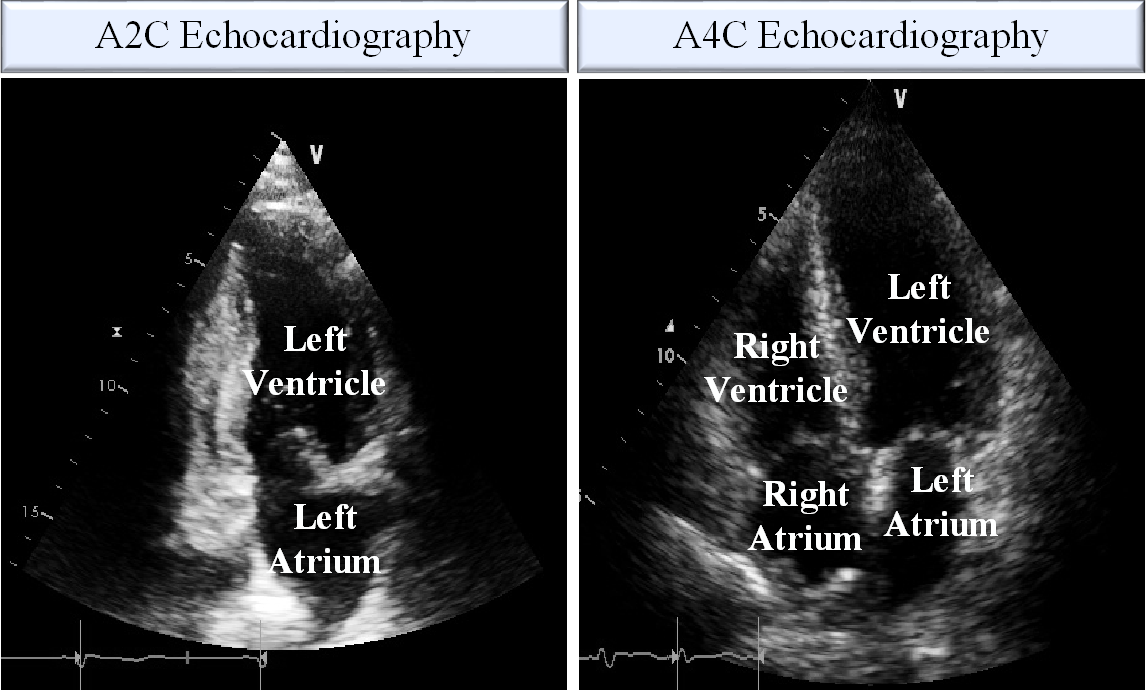}
    \caption{The chambers of the heart in A2C and A4C view echocardiography.}
    \label{fig:chambers}
\end{figure}

Gradient-based optical flow methods estimate the motion by capturing the flow of pixels with constant intensity over time. The optical flow can be described as the velocity distributions of the bright pixel movements in the image that can be approximated by the partial derivatives with respect to spatial and temporal coordinates. Several studies \cite{6341835, BEHAR2004743, 1407980, lucas1981iterative} have utilized gradient-based optical flow methods by adding constraints to regularize the myocardial motion in 2D echocardiography. However, the ultrasound imaging is highly variant with the angle and depth of the ultrasound beam, which results in many artifacts, such as noise, shadowing, and dropouts on the image that cause optical flow methods to fail at estimating large LV wall displacements \cite{1175086, 8039230}. Moreover, the noisy nature of echocardiography degrades the performance of the optical flow algorithms. 

Deformation imaging is widely used in echocardiography to perform strain analysis of the myocardium \cite{1501918, BEHAR200457, AMUNDSEN2006789, 6189174, 8760511, bansal2010assessment, jamal2001noninvasive, leitman2004two, riffel2015assessment, shah2012myocardial}. The strain is defined by the length of the LV wall that is measured via the speckle tracking method, which is known also as the region-based optical flow or block-matching method in echocardiography. The speckle tracking method searches for a similar block of pixels through consecutive frames in a specified search window. The strain measurements refer to deformation of the myocardium, e.g., if the velocity of the LV wall segments is non-uniform, then the myocardium is infarcted. Even though deformation imaging is a promising method to detect MI, it suffers severely from the weakness of the optical flow methods, which are not robust to the noisy nature of echocardiography. Thus, there is a need for high-quality echocardiography recordings with $50-70$ frames per second (fps) in order to tackle the issues raised by the speckle tracking in deformation imaging \cite{dandel2009strain}. As a result, the clinical usage of deformation imaging is limited.

The active contour (snake) is introduced by Kass \textit{et al.} \cite{kass1988snakes} that evolves iteratively to minimize the energy curves to extract the edges, lines, or boundaries in images. They are used in studies \cite{996341, 500138, landgren2013segmentation} to extract and track the endocardial boundary of the LV wall in echocardiography. However, the endocardial boundary is often discontinuous in echocardiography due to the high level of noise. These occlusions and indentations on the LV wall cause snake to fail at extracting to the true boundary \cite{9354781}. 

In this study, in order to overcome the aforementioned limitations of the motion estimation algorithms, we use the Active Polynomials (APs) \cite{9261387} that constrain the active contours to achieve robust segmentation and tracking of the endocardial boundary of the LV wall. In our previous study \cite{9261387}, we proposed a single-view MI detection approach by thresholding the maximum displacement of APs in A4C view echocardiography. However, setting a fixed threshold is never guaranteed to be optimal for decision-making. Moreover, in the literature, many studies have also used single-view, high-quality, or simulated echocardiographic data \cite{996341}. Additionally, in cardiology, several studies \cite{kusunose2020deep, omar2018automated, vidya2015computer} diagnose MI using conventional and Deep Learning methods. However, one major limitation is that they require large datasets for training. Therefore, the reliability and performance of the previously proposed methods may significantly vary as the clinical data is usually scarce, in low quality/resolution, and subject to a high level of noise. Hence, our first objective in this study is to improve the performance of our previous work \cite{9261387} by bringing an intelligent diagnosis via Machine Learning over the hand-crafted features to surpass threshold-based diagnosis in single-view echocardiography.

A common and major drawback of all the prior studies in this domain is that the proposed MI detection methods all rely on single-view, mostly over the apical 4-chamber view. Only certain segments can be analyzed on a single-view and this brings an inevitable problem of missing the ongoing MI if the RWMA is not present on those segments. In other words, regardless of their accuracy, if the segment(s) that show the abnormal motion is not visible on that particular echocardiography view, they are bound to fail the MI detection. Therefore, the second objective of this study is to diagnose MI on the LV wall by using multi-view echocardiography, which includes apical 4-chamber (A4C) and apical 2-chamber (A2C) views, where all the chambers of the heart, and only left atrium and LV are visible, respectively  as it is depicted in Fig. \ref{fig:chambers}. 

\begin{figure*}[t!]
    \centering
    \includegraphics[width=1\linewidth]{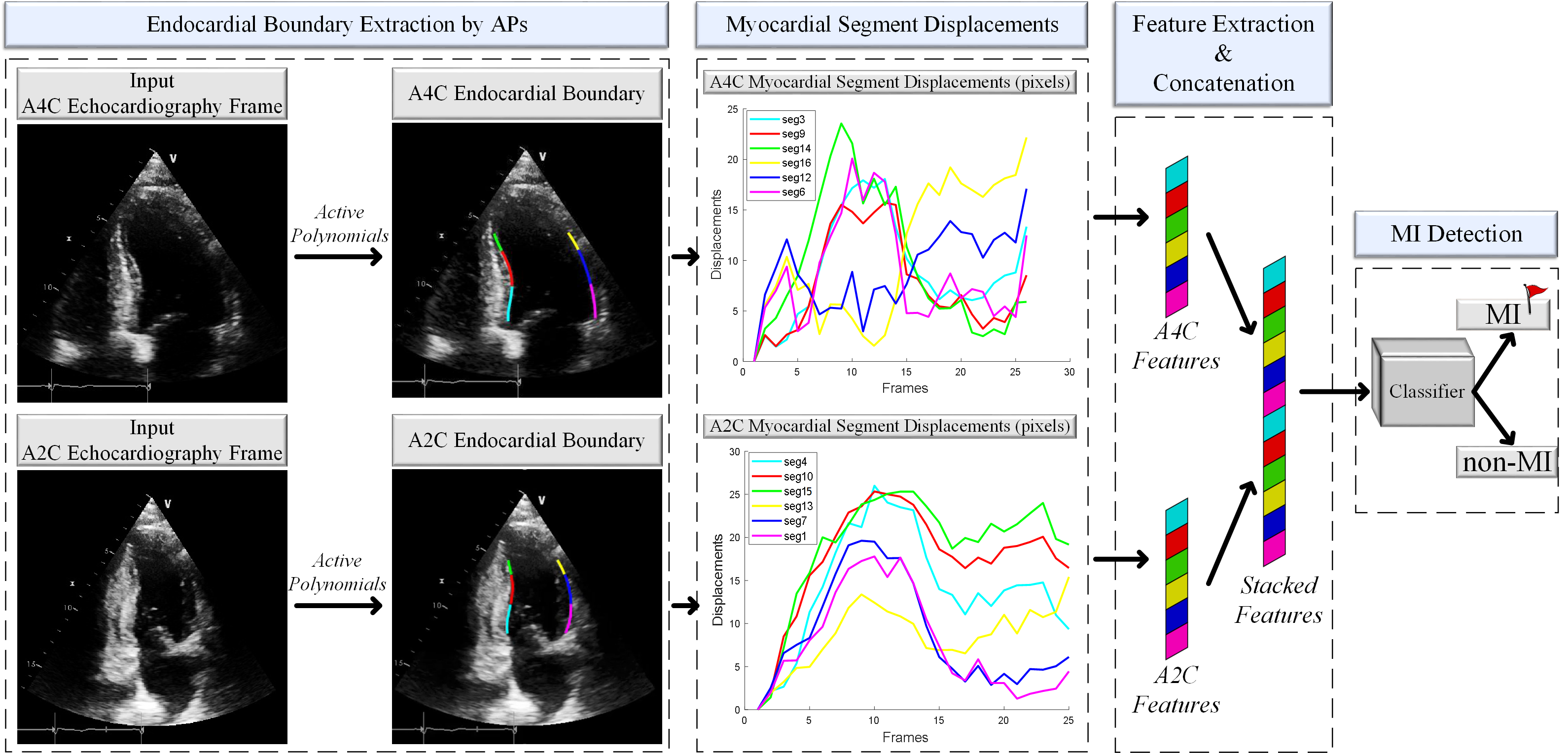}
    \caption{The diagram of the proposed MI detection approach using multi-view echocardiography. The endocardial boundary is first extracted by the APs method. Then, the defined myocardial segments are tracked through one-cardiac cycle to form the displacement curves. The maximum displacements are generated from each segment to define the features that are then concatenated and fed into the classifier to detect MI.}
    \label{fig:method}
\end{figure*}

In order to improve robustness and generalization of MI diagnosis, in this study, we propose a multi-view Machine Learning (ML) approach over the maximum displacement features as depicted in Fig. \ref{fig:method}. As the pioneer MI diagnosis study in the literature over multi-view echocardiography, we aim to determine the best ML approach for this purpose. Therefore, we perform an extensive set of comparative evaluations among several ML methods including Decision Tree (DT), Random Forest (RF), k-Nearest Neighbour (k-NN), Support Vector Machine (SVM), and 1D-Convolutional Neural Networks (1D-CNN). Accordingly, the contributions of this study are summarized as follows:
\begin{itemize}
    \item We improve the performance of threshold-based APs \cite{9261387} over single-view echocardiography. Additionally, APs are adapted for the A2C view echocardiography for the first time. 
    \item We propose a pioneer algorithm for MI diagnosis using multi-view echocardiography. Noting that the ground-truth labels of single-view and multi-view echocardiography are different, direct comparison is not viable. Hence, our study reveals the results of multi-view echocardiography for the first time to perform a \textit{reliable} MI diagnosis by considering more myocardial segments in the analysis phase.
    \item An extended benchmark dataset, HMC-QU\footnote{The benchmark HMC-QU dataset is publicly shared with the research community at the repository \href{https://www.kaggle.com/aysendegerli/hmcqu-dataset}{https://www.kaggle.com/aysendegerli/hmcqu-dataset}.}  is created that includes $260$ echocardiography recordings of $130$ MI patients and healthy subjects from A4C and A2C views.
\end{itemize}

The rest of the article is organized as follows. In Section \ref{sec:methods}, we give the details of the proposed approach. In Section \ref{sec:dataset}, we introduce the HMC-QU dataset, and in Sections \ref{sec:experimental-results-single} and \ref{sec:experimental-results-multi} we report the experimental results. Finally, we conclude the paper and suggest topics for future research in Section \ref{sec:conclusion}.

\section{Methodology}\label{sec:methods}
In this section, the proposed approach will be described in detail. As it is depicted in Fig. \ref{fig:method}, in the first step, the endocardial boundary of the LV wall is extracted by APs. Then, the boundary is divided into myocardial segments from which the displacement curves are generated. Lastly, the features are extracted from the displacements of each myocardial segment, which are then used as the input for the classifiers for MI diagnosis. 

\subsection{Endocardial Boundary Extraction by Active Polynomials}
\begin{figure*}[t!]
    \centering
    \includegraphics[width=0.85\linewidth]{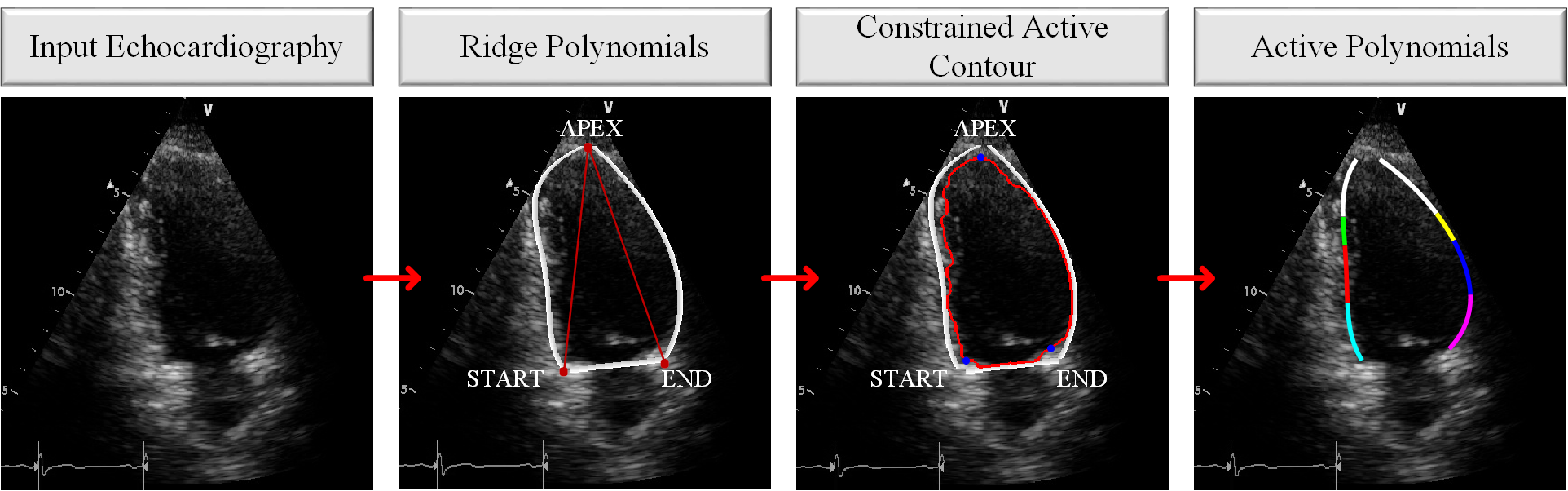}
    \caption{The APs method for the endocardial boundary of the LV wall extraction consists of three stages: 1) the RPs on the LV wall are formed in input echocardiography, 2) the active contour is evolved from inside of the LV and constrained by the RPs, and 3) the APs are formed by fitting $4^{\text{th}}-$order polynomials on the evolved active contour.}
    \label{fig:APs}
\end{figure*}

Accurate extraction of the LV wall is crucial to obtain the true motion of the myocardium. In order to overcome the limitations of the active contours \cite{kass1988snakes} in echocardiography, we use Active Polynomials \cite{9261387} to extract the endocardial boundary of the LV wall. Echocardiography is usually subjected to a high level of noise, and during acquisition, some parts of the chamber walls might be missing or out of view. APs provide a robust and reliable segmentation and tracking of the LV wall, where their formation is illustrated in Fig. \ref{fig:APs}. A brief summary will be presented next and the details of the method can be found in \cite{9261387}.

\begin{figure}[b!]
    \centering
    \includegraphics[width=\linewidth]{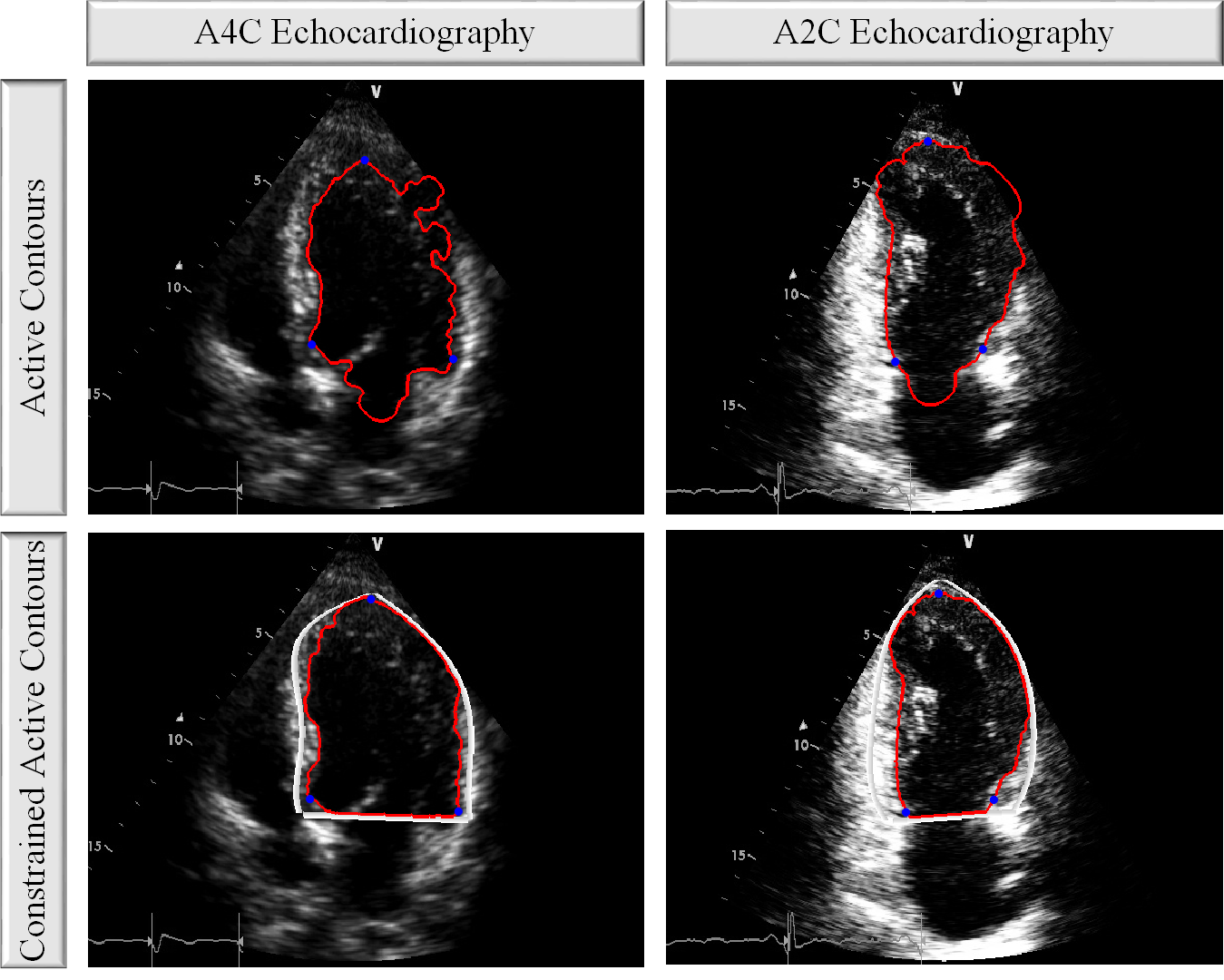}
    \caption{The comparison of the original and constrained active contours. In both A4C and A2C views, the constrained active contours can extract the endocardial boundary more accurately.}
    \label{fig:constrained}
\end{figure}

\begin{figure}[b!]
    \centering
    \includegraphics[width=\linewidth]{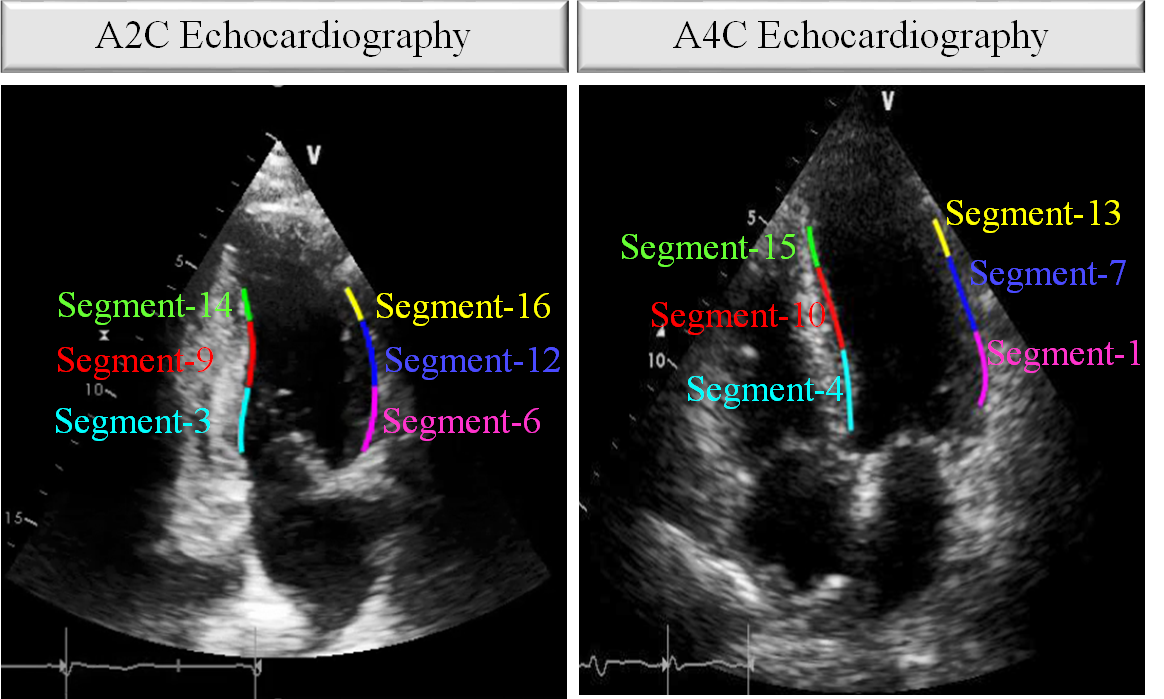}
    \caption{The myocardial segments of A4C and A2C views echocardiography based on the $17-$segment model.}
    \label{fig:myocardial_Segments}
\end{figure}

In the first stage of the APs formation, the Ridge Polynomials (RPs) are formed on the LV wall. In echocardiography, the LV wall may partially be missing or invisible due to low quality. Thus, in the evolvement process of the active contours, the contour may escape from the chamber causing inaccurate segmentation of the endocardial boundary. Therefore, the RPs are first created to constrain the active contours as illustrated in Fig. \ref{fig:constrained}. In the second stage of the proposed method, we initialize an active contour from inside the chamber. The initial mask for the contour is located in the middle of the LV as a mini-version of the current frame’s RPs. The aim is to evolve an active contour to detect and extract the endocardial boundary of the LV wall. A typical edge detector is expressed as follows:
\begin{equation}
    \lim_{z\to\infty} g(z)=0,
    \label{eq1}
\end{equation}
\noindent where $g$ is a function with positive and decreasing values, $z$ is an image, and the edges of $z$ are detected at the locations where the gradient is zero. However, detecting the edges of images with rough and discontinuous objects is challenging with the gradient method since generally gradient is not zero on that particular edges. Thus, Chan-Vese \cite{chan2001active} active contour method is utilized since its stopping criteria do not depend on the gradient. Therefore, it is suitable especially for echocardiography, where there are discontinuities (even though it has been minimized by RPs) and rough edges on the LV wall due to the high level of noise and acquisition. Once the active contour has converged to the endocardial boundary, the APs can then be formed over the evolved active contour. As shown in Fig. \ref{fig:constrained}, the active contour may be noisy with severe discontinuities on the LV wall. In order to achieve a smooth endocardial boundary segmentation, the evolved active contour is divided into two sections. The left part of the contour corresponds to the active contour points from start to apex, whereas the right part is from apex to end. After the division, we compose $4^{\text{th}}-$order smooth polynomials each of which is fitted to the equally distanced $9$ points from both right and left parts to form APs that are the final form of the endocardial boundary. Thus, APs provide a robust and smooth endocardial boundary for MI diagnosis.

\begin{figure}[b!]
    \centering
    \includegraphics[width=\linewidth]{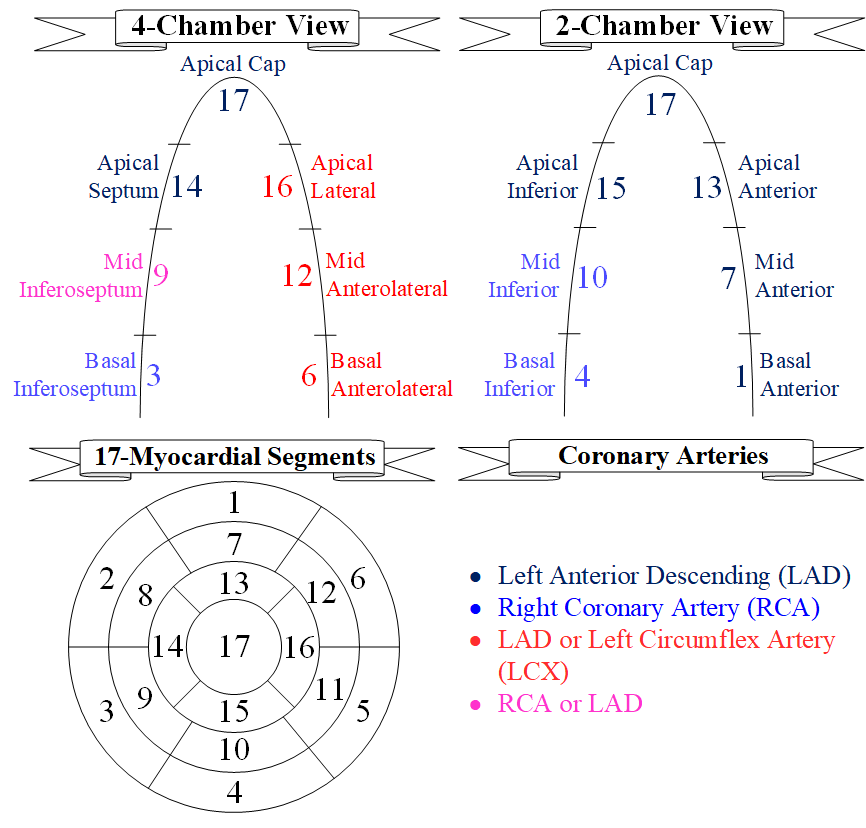}
    \caption{The myocardial segment names and numbers are shown for both A4C and A2C views echocardiography at the top row. The bull eye's plot of the $17-$segment model is illustrated, where each color-coded and numbered segment corresponds to coronary arteries at the bottom row.}
    \label{fig:segment_diagram}
\end{figure}

\begin{figure*}[t!]
    \centering
    \includegraphics[width=0.8\linewidth]{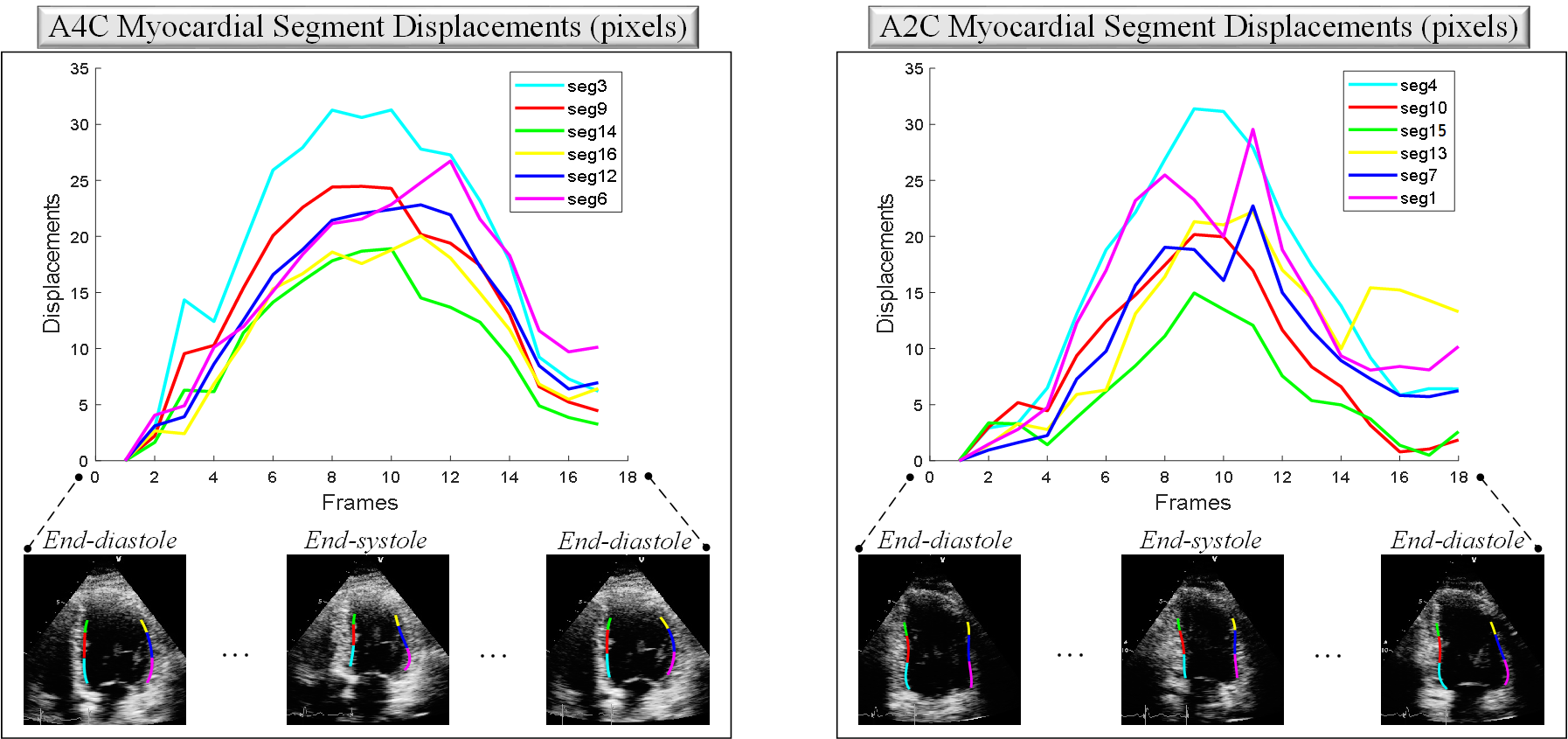}
    \caption{The displacement curves of A4C and A2C view echocardiography recordings of a patient, where the frames consist of one-cardiac cycle.}
    \label{fig:disp_curves}
\end{figure*}

\subsection{Myocardial Segment Displacements}
After extracting the endocardial boundary by APs in each frame of the echocardiography recordings, the boundary is tracked and its displacement is measured in one-cardiac cycle. In the diagnosis, the LV wall is segmented into $17$ myocardial segments, which is the recommendation of the American Heart Association Writing Group on Myocardial Segmentation and Registration for Cardiac Imaging \cite{lang2005recommendations}. The myocardial segments on the LV wall for both A4C and A2C views can be seen in Fig. \ref{fig:myocardial_Segments} with a total of $12$ distinct myocardial segments. It is recommended that segment$-17$ should be removed if the wall motion or regional strain are analyzed using the $17-$segment model \cite{lang2015recommendations}. Thus, in the analysis, we have excluded segment$-17$ shown as the white myocardial segment in the APs block in Fig. \ref{fig:APs}. Consequently, analyzing the motion of the $12$ myocardial segments in multi-view echocardiography yields information regarding all the coronary arteries feeding the heart muscle since they cover most of the heart area as illustrated in Fig. \ref{fig:segment_diagram}. The myocardial segments are divided based on the length of the APs that are formed at the end of the endocardial boundary extraction process. The length of the APs are considered individually as previously explained left and right parts, where the length of the left part is represented as $L$, whereas the right part's length is $R$. Accordingly, in A2C and A4C views, the length of apical myocardial segments are  $R/7$ and $L/7$, whereas other segments have $2R/7$ and $2L/7$ for right and left parts, respectively.

The displacement curves are plotted for each myocardial segment in A2C and A4C views echocardiography. The displacements are calculated over one-cardiac cycle echocardiography recordings, where the reference frame is at time $t=0$. At each time instance, the displacement is defined as follows:
\begin{equation}
    \begin{split}
     D_{s_{\kappa}}(t) = & \frac{1}{N}\sum_{i=1}^N \\\sqrt{(x_{s_\kappa}^i(t=0) - x_{s_\kappa}^i(t))^2 + (y_{s_\kappa}^i(t=0) - y_{s_\kappa}^i(t))^2}, 
     \end{split}
    \label{eq8}
\end{equation}
\noindent where $D_{s_\kappa}$ is the average displacement measure of the $\kappa$ numbered myocardial segment $s_{\kappa}$ at a time instant $t$, $N$ is the number of points equally taken on the myocardial segment, i.e., $N=5$ in our implementation, and $(x, y)$ is the coordinate of each point taken. Accordingly, at the reference frame, the displacement measurement is equal to zero. It is expected that the displacement measures of each segment would gradually increase from end-diastole to middle of the cycle; on the other hand, gradually decrease from middle of the cycle to end-systole as illustrated in Fig. \ref{fig:disp_curves}. 

\subsection{Feature Engineering}
In the feature engineering stage, we extract information from the displacement curves by quantitatively imitating the evaluation of cardiologists. A cardiologist visually assesses the RWMA by correlating the infarction to the displacement measurement of a myocardial segment. Accordingly, the larger the displacement measure a myocardial segment has, the less the chance of it being infarcted. Thus, the maximum displacement of each myocardial segment is extracted as the features. However, a myocardial segment displacement cannot directly be compared since the displacements decrease gradually from valve to apical cap due to the structure of the heart. Therefore, the displacements of the myocardial segments are normalized by dividing the maximum displacement of a segment with the minimum interval between the segment and the other segment at the opposite side. For example, the interval between segment$-3$ and its opposite segment$-6$ is greater than the interval between segment$-14$ and its opposite segment$-16$. Thus, we bring each displacement to the same level for a fair comparison. The interval measurement used in the normalization is defined as follows:
\begin{equation}
    I_{(s_\kappa, s_\varepsilon)}(t) = \frac{1}{N}\sum_{i=1}^N |x_{s_\kappa}^i(t) - x_{s_\varepsilon}^i(t)| + |y_{s_\kappa}^i(t) - y_{s_\varepsilon}^i(t)|,
    \label{eq9}
\end{equation}
\noindent where $I_{(s_\kappa, s_\varepsilon)}$ is the averaged Manhattan distance of $N=5$ number of equally taken points on the two opposite segments $s_\kappa \text{, } s_\varepsilon$  at time $t$, and $\kappa \text{, } \varepsilon$ are the segment numbers. Accordingly, we form the features of each myocardial segment as follows:
\begin{equation}
    f_{s_\kappa}=\frac{max(\mathbf{D_{s_\kappa})}}{min(\mathbf{I_{(s_\kappa, s_\varepsilon)})}},
    \label{eq10}
\end{equation}
\noindent where $f$ is the displacement feature of segment $s_\kappa$ that is the maximum displacement divided by the minimum interval between its opposite segment $s_\varepsilon$. For the displacement calculation, we have used the Euclidean distance, whereas, for the interval measurement, the Manhattan distance is utilized to scale the features into $[0, 1]$ more precisely as adapted by the threshold-based APs \cite{9261387} approach. Consequently, in single-view echocardiography, where we only use whether A4C or A2C view echocardiography recording, we extract feature vectors $\mathbf{\Phi_{1}} \text{, } \mathbf{\Phi_{2}} \in \mathbb{R}^{6\times1}$ in one-cardiac cycle, respectively defined as follows: 
\begin{equation}
    \begin{aligned}
    \mathbf{\Phi_1} = \begin{bmatrix} 
                f_{s_3}\\
                f_{s_9}\\
                f_{s_{14}}\\
                f_{s_{16}}\\
                f_{s_{12}}\\
                f_{s_{6}}\\
    \end{bmatrix} \text{, and }
    \mathbf{\Phi_2} = \begin{bmatrix} 
                f_{s_4}\\
                f_{s_{10}}\\
                f_{s_{15}}\\
                f_{s_{13}}\\
                f_{s_7}\\
                f_{s_1}\\
    \end{bmatrix},
    \end{aligned}
    \label{eq11}
\end{equation}
\noindent where $s$ denotes the numbered myocardial segment features as illustrated in Fig. \ref{fig:segment_diagram} and calculated in Eq. (\ref{eq10}). On the other hand, in multi-view echocardiography, we concatenate the feature vectors to form $\mathbf{F} = \begin{bmatrix} \mathbf{\Phi_1}^\text{T} \text{ } \mathbf{\Phi_2}^\text{T} \end{bmatrix} \in \mathbb{R}^{1\times12}$.

\subsection{MI Detection}
The MI detection is performed via binary classification task, where the extracted features are fed into several classifiers as follows: Support Vector Machine (SVM), k-Nearest Neighbors (k-NN), Decision Tree (DT), Random Forest (RF), and 1D-Convolutional Neural Networks (1D-CNN). The training is performed over $K$ number of samples $\{f_{train}^j, L_{train}^j\}^K_{j=1}$, where $f$ and $L$ are the data and ground-truths, respectively. 

\textbf{Support Vector Machines.} The binary classification task via SVM is performed by separating the data with a hyperplane \cite{cristianini2000introduction}. The best-fitting hyperplane is determined by maximizing the inter-class and minimizing the intra-class differences. In order to impose non-linearity to SVM models, kernel-based methods are used to construct non-linear features that map the data into higher dimensions to perform an easier class separation. Thus, the performance of classification can be improved and the overfitting can be avoided.

\textbf{Decision Tree.} The hierarchical structure of DT performs a classification task by feeding the data to nodes that are divided into branches for transferring the input to the most suitable class label \cite{breiman1984ra}. The tree is formed by selecting the nodes as they are divided into branches, and whenever the stopping criterion is satisfied, the final node is assigned to a class. DT models are suitable for small datasets, and computationally less expensive compared to other models used in this study. 

\textbf{Random Forest.} As an ensemble version of the DT, the RF prevents the overfitting issue that occurs due to the tight-fitting of the model to the training data. Overfitting causes the generalization capability of the model to deteriorate. Therefore, the RF model overcomes the overfitting issue by constructing individual trees by minimizing their correlation in the classification task. After the majority voting, the best model is selected to be used for the task. 

\textbf{k-Nearest Neighbors.} The k-NN method classifies data by assigning a sample to the same class as its k-nearest neighbours \cite{gowda1979condensed}. It is popular due to its robustness to noise and the simplicity of the algorithm. Moreover, it requires a few parameters to tune, which makes the cross-validation process straightforward \cite{goldberger2004neighbourhood}. The performance of the k-NN method improves as more data is used to train it. However, more training data increases its computational cost and memory consumption since k-NN stores the training data in order to calculate the distance between samples to classify a test sample.  

\begin{figure}[t!]
    \centering
    \includegraphics[width=\linewidth]{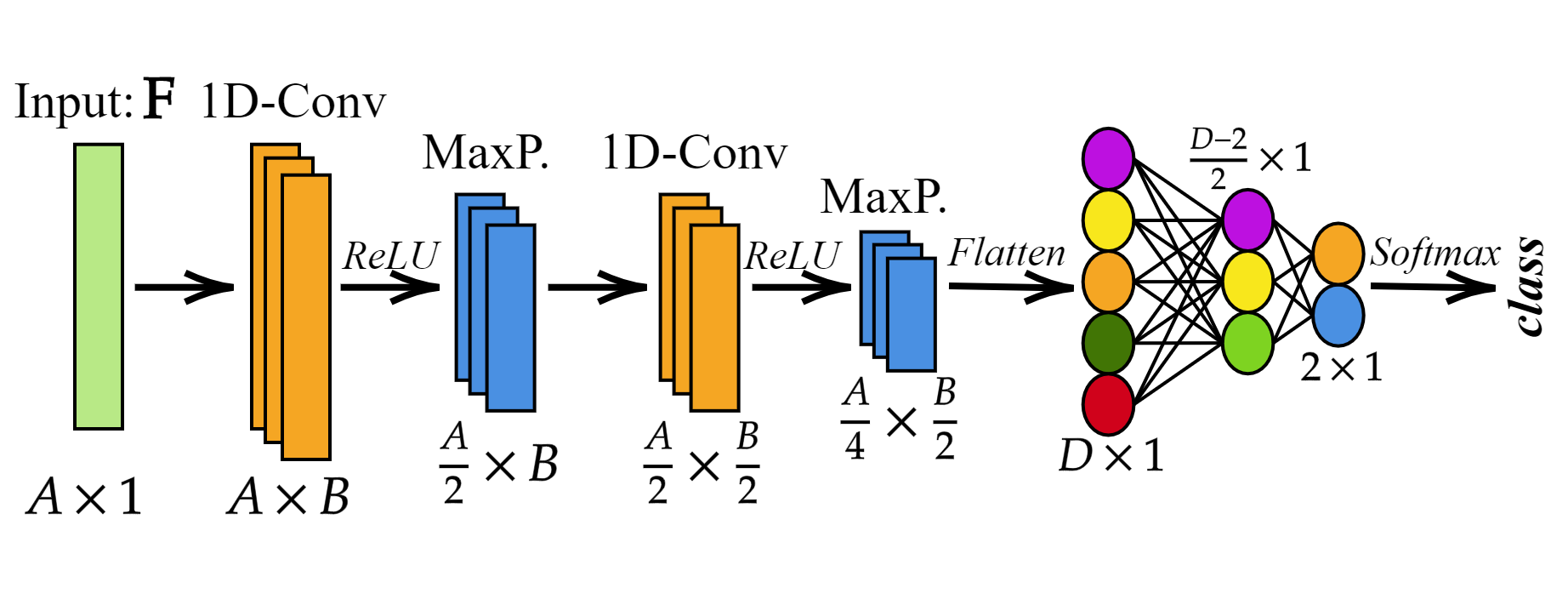}
    \caption{The proposed 1D-CNN structure consists of two 1D-convolutional (1D-Conv) and two max-pooling (MaxP.) layers. The input, filter, and fully connected layer sizes are denoted as $A$, $B$, and $D$, respectively.}
    \label{fig:CNN}
\end{figure}

\textbf{1D-Convolutional Neural Networks.} The most popular ML method during the last decade is Convolutional Neural Networks (CNNs) that are feed-forward models consisting of input, output, and hidden layers \cite{KIRANYAZ2021107398}. Their difference to Artificial Neural Networks is that convolution operations are performed in the hidden layers. In one-dimensional signal processing applications, 1D-CNNs are preferred due to their feasibility to one-dimensional convolution operations and low computational complexity compared to 2D-CNNs \cite{KIRANYAZ2021107398}. The 1D-CNN model proposed in this study maps the input feature, $\mathbf{F}$ to the corresponding class label, $\mathbf{L}:\mathbf{L}\xleftarrow[]{}P_{\vartheta, \chi}(\mathbf{F})$. The first block $\vartheta \in \{b_j, w_j\}^M_{j=1}$ consists of $M=2$ number of 1D-convolutional layers, Rectified Linear Unit (ReLU) activation function, and max-pooling layers by the size of $2$, respectively. The second block $\chi$ consists of a fully connected layer, ReLU activation function, an output layer, and softmax activation function, respectively. The filter and kernel sizes of the convolutional layers are presented in Section \ref{sec:experimental-setup}. Accordingly, the block diagram of the proposed 1D-CNN is illustrated in Fig. \ref{fig:CNN}, where its compact structure avoids overfitting.

\section{Experimental Results}\label{sec:experiments}
In this section, we detail the HMC-QU dataset and report the experimental results for single-view and multi-view echocardiography.

\subsection{HMC-QU Dataset}\label{sec:dataset}
The cardiologists of Hamad Medical Corporation, and researchers from Qatar University and Tampere University have compiled the HMC-QU dataset that includes 2D echocardiography recordings for MI detection. This benchmark dataset has been approved for usage by the local ethics board of the hospital in February $2019$. The dataset consists of $260$ recordings from A2C and A4C views of $130$ subjects. The MI term indicates any sign of RWMA, whereas subjects without RMWA are labeled as non-MI in the dataset. All the MI patients had first-time acute MI and were admitted with ECG and cardiac enzymes evidence. The patients were treated with coronary angiography/angioplasty whose echocardiography recordings are taken within $24$ hours after admission or before the operation. Non-MI subjects were not diagnosed with MI according to their echocardiography recordings, but they underwent a required health check in the hospital.

\begin{table}[t!]
\centering
\caption{The number of myocardial segments corresponds to the ground-truth labels of the HMC-QU dataset.}
\resizebox{.43\textwidth}{!}{
\begin{tabular}{ccc}
\hline\hline
Myocardial Segments & MI Patients & non-MI Subjects \\ \hline\hline
Segment$-1$           & $29$          & $101$             \\ 
\rowcolor[gray]{.97} Segment$-3$      & $26$    & $104$   \\ 
Segment$-4$           & $29$          & $101$             \\ 
\rowcolor[gray]{.97}Segment$-6$       & $16$    & $114$   \\ 
Segment$-7$           & $40$          & $90$              \\
\rowcolor[gray]{.97}Segment$-9$       & $46$    & $84$    \\
Segment$-10$          & $31$          & $99$              \\ 
\rowcolor[gray]{.97}Segment$-12$      & $28$    & $102$    \\ 
Segment$-13$          & $47$          & $83$              \\ 
\rowcolor[gray]{.97}Segment$-14$      & $64$    & $66$    \\ 
Segment$-15$          & $53$          & $77$      \\ 
\rowcolor[gray]{.97}Segment$-16$      & $53$    & $77$    \\ 
\end{tabular}}
\label{tab:segment_numbers}
\end{table}

The number of myocardial segments with respect to ground-truth labels are presented in Table \ref{tab:segment_numbers}. The A4C view includes $80$ MI and $50$ non-MI recordings, whereas $68$ MI and $62$ non-MI recordings are from the A2C view. Accordingly, MI ratios are $61.54\%$ and $52.3\%$ in A4C and A2C views, respectively. The ratios differ from each other since only $60$ patients have their MI visible in both views. Therefore, in multi-view echocardiography, the ground-truth labels correspond to $88$ MI patients and $42$ non-MI subjects, where the ground-truth labels are formed as MI if \textit{any} of the views depict RWMA, whereas non-MI if no sign of RWMA is visible in both views. Thus, the overall MI ratio is $67.69\%$ in multi-view echocardiography. In Table \ref{tab:MI_ratios}, a detailed ground-truth label formation with respect to views is presented. 

\begin{table}[b!]
\centering
\caption{The number of patients with respect to their corresponding ground-truth labels from A4C and A2C views.}
\resizebox{.34\textwidth}{!}{
\begin{tabular}{c|c|c}
\hline
\multicolumn{2}{c|}{Ground-truths} & \multirow{2}{*}{\# of Patients} \\ \cline{1-2}
A4C view & A2C view &  \\ \hline 
MI & MI & 60 \\
\rowcolor[gray]{.97}non-MI & non-MI & 42 \\
MI & non-MI & 20 \\
\rowcolor[gray]{.97}non-MI & MI & 8 \\ \hline
\end{tabular}}
\label{tab:MI_ratios}
\end{table}

In each echocardiography recording, the myocardial segments on the LV wall are categorized into five different stages: $1-$normal or hyperkinesia, $2-$hypokinesia, $3-$akinesia, $4-$dyskinesia, and $5-$aneurysm as the severity of MI ascends, respectively. In this study, we perform a binary classification task to simplify the problem. Therefore, we have downsized the ground-truth labels to $1-$non-MI (normal), and $(2, 3, 4, 5)-$MI. The ultrasound machines used for acquisition are Phillips and GE Vivid from GE Healthcare (United States). The spatial resolution of the echocardiography recordings varies from $(422\times636)$ to $(768\times1024)$, and the temporal resolution is $25$ frames per second (fps).

\subsection{Experimental Setup}\label{sec:experimental-setup}
The detection models are evaluated over the dataset in a stratified $5$-fold cross-validation scheme with a ratio of 80\% training, and 20\% test (unseen data) sets considering a balanced ratio of classes. The confusion matrices are formed by the elements: true positive ($TP$), true negative ($TN$), false positive ($FP$), and false negative ($FN$). Thus, the standard performance metrics are calculated as follows:
\begin{equation}
    Sensitivity = \frac{TP}{TP+FN},
\end{equation}
\noindent where the sensitivity (recall) is the ratio of correctly detected MI patients to all MI patients in the dataset,
\begin{equation}
    Specificity = \frac{TN}{TN+FP},
\end{equation}
\noindent where the specificity is the ratio of correctly classified non-MI subjects to all non-MI subjects in the dataset, 
\begin{equation}
    Precision = \frac{TP}{TP+FP},
\end{equation}
\noindent where the precision refers to the number of correctly detected MI patients over the total number of samples detected as positive class in the dataset,
\begin{equation}
    Accuracy = \frac{TP+TN}{TP+TN+FP+FN},
\end{equation}
\noindent where the accuracy is the ratio of correctly detected samples in the dataset,
\begin{equation}
    F(\beta) = (1+\beta^2)\frac{Precision \times Sensitivity}{\beta^2 \times Precision + Sensitivity},
\end{equation}
\noindent where the F$1-$Score and F$2-$Score are calculated as the weighting parameter $\beta=1$ and $\beta=2$, respectively. The F$1-$Score refers to the harmonic average of precision and sensitivity metrics. On the other hand, F$2-$Score emphasizes the sensitivity metric with a higher $\beta$ value. Consequently, the objective of the detection phase is to maximize sensitivity with a preferable specificity to avoid missing MI patients. Moreover, F$2-$Score is targeted to be maximized with a reasonable F$1-$Score value.

\begin{table*}[t!]
\centering
\caption{Average MI detection performance results (\%) computed from 5-folds in single-view echocardiography.}
\resizebox{.85\linewidth}{!}{
\begin{tabular}{ccccccccc}
\hline \hline
& MI Ratios & Model & \multicolumn{1}{c}{Sensitivity} & \multicolumn{1}{c}{Specificity} & \multicolumn{1}{c}{Precision} & \multicolumn{1}{c}{F$1-$Score} & \multicolumn{1}{c}{F$2-$Score} & \multicolumn{1}{c}{Accuracy} \\ 
\hline \hline

\multirow{6}{*}{\rotatebox[]{90}{\large{A4C}}} & \multirow{6}{*}{$61.54\%$}& \cellcolor[gray]{.97}DT & \multicolumn{1}{c}{\cellcolor[gray]{.97}$85.00$} & \multicolumn{1}{c}{\cellcolor[gray]{.97}$68.00$} & \multicolumn{1}{c}{\cellcolor[gray]{.97}$80.95$} & \multicolumn{1}{c}{\cellcolor[gray]{.97}$82.93$} & \multicolumn{1}{c}{\cellcolor[gray]{.97}$84.16$} & \multicolumn{1}{c}{\cellcolor[gray]{.97}$78.46$} \\

 & & RF & \multicolumn{1}{c}{$86.25$} & \multicolumn{1}{c}{$\textbf{84.00}$} & \multicolumn{1}{c}{$\textbf{89.61}$} & \multicolumn{1}{c}{$\textbf{87.90}$} & \multicolumn{1}{c}{$86.90$} & \multicolumn{1}{c}{$\textbf{85.38}$} \\
 
 & & \cellcolor[gray]{.97}SVM & \multicolumn{1}{c}{\cellcolor[gray]{.97}$82.50$} & \multicolumn{1}{c}{\cellcolor[gray]{.97}$70.00$} & \multicolumn{1}{c}{\cellcolor[gray]{.97}$81.48$} & \multicolumn{1}{c}{\cellcolor[gray]{.97}$81.99$} & \multicolumn{1}{c}{\cellcolor[gray]{.97}$82.29$} & \multicolumn{1}{c}{\cellcolor[gray]{.97}$77.69$}  \\
 
 & & \multicolumn{1}{c}{k-NN} & $\textbf{88.75}$ & $78.00$ & $86.59$ & $87.65$ & $\textbf{88.31}$ & $84.62$ \\
 
 & & \multicolumn{1}{c}{\cellcolor[gray]{.97}1D-CNN} & \cellcolor[gray]{.97}$83.75$ & \cellcolor[gray]{.97}$78.00$ & \cellcolor[gray]{.97}$85.90$ & \cellcolor[gray]{.97}$84.81$ & \cellcolor[gray]{.97}$84.17$ & \cellcolor[gray]{.97}$81.54$  \\ 
 
 & & \multicolumn{1}{c}{APs\cite{9261387}} & $86.25$ & $77.08$ & $86.25$ & $86.25$ & $86.25$ & $82.81$ \\
 \hline
 
 \multirow{6}{*}{\rotatebox[]{90}{\large{A2C}}} &\multirow{6}{*}{$52.30\%$}& \cellcolor[gray]{.97}DT & \cellcolor[gray]{.97}$67.65$ & \cellcolor[gray]{.97}$58.06$ & \cellcolor[gray]{.97}$63.89$ & \cellcolor[gray]{.97}$65.71$ & \cellcolor[gray]{.97}$66.86$ & \cellcolor[gray]{.97}$63.08$ \\
 
 & & RF & $66.18$ & $\textbf{77.42}$ & $76.27$  & $70.87$ & $67.98$  & $71.54$ \\
 
 & & \cellcolor[gray]{.97}SVM & \cellcolor[gray]{.97}$\textbf{76.47}$ & \cellcolor[gray]{.97}$66.13$ & \cellcolor[gray]{.97}$71.23$ & \cellcolor[gray]{.97}$73.76$ & \cellcolor[gray]{.97}$\textbf{75.36}$ & \cellcolor[gray]{.97}$71.54$  \\
 
 & & \multicolumn{1}{c}{k-NN} & $72.06$  & $\textbf{77.42}$  & $\textbf{77.78}$  & $\textbf{74.81}$  & $73.13$  & $\textbf{74.62}$  \\
 
 & & \multicolumn{1}{c}{\cellcolor[gray]{.97}1D-CNN} & \cellcolor[gray]{.97}$64.71$ & \cellcolor[gray]{.97}$75.81$ & \cellcolor[gray]{.97}$74.58$ & \cellcolor[gray]{.97}$69.29$ & \cellcolor[gray]{.97}$66.47$ & \cellcolor[gray]{.97}$70.00$ \\
 
  & & \multicolumn{1}{c}{APs\cite{9261387}} & $69.12$ & $59.68 $ & $65.28$ & $67.14$ & $68.31$ & $64.62$ \\
 \hline
\end{tabular}}
\label{tab:experiments_1}
\end{table*}

The implementation of the detection models is performed on Python using the TensorFlow library \cite{abadi2016tensorflow} and Scikit-learn library \cite{scikit-learn}, whereas the feature engineering of the proposed method is implemented on MATLAB version R$2019$a. For the experiments, we have used a PC with Intel® i$7-8665$U CPU $32$ GB system memory, and a workstation with NVidia® GeForce RTX 2080 Ti GPU card $128$ GB system memory. In the training phase of each classifier, we have performed a grid search over a $5$-fold cross-validation scheme that is an exhaustive search of specified parameter values for each model in order to set the best hyper-parameters for the testing phase. The grid search is performed by the \textit{GridSearchCV} function of the Scikit-learn library. The best parameters are selected according to the highest F$1-$Score over the validation data, which is extracted from the training set of each fold. Thus, in the testing phase, the best parameters are set for each fold differently. Accordingly, we search the best parameters of the classifiers as follows: 

\textbf{DT} has searched the function of \textit{Gini impurity} and \textit{entropy} for measuring the quality of a split, the maximum number of features that are selected for the best split is defined by the \textit{auto},  \textit{$log2$}, and \textit{square root} of the number of features in the training set, the nodes are separated by the supported strategies that are set to \textit{random} and \textit{best}, and the performance of the model is evaluated on the test set by checking the scoring of each standard performance metrics.

\textbf{RF} classifier has bootstrap parameter set to \textit{false} and \textit{true} that indicates the data usage as building the trees, the class weights are determined by \textit{balanced} and \textit{balanced subsample} mode, the quality of splits are measured by \textit{Gini impurity} and \textit{entropy} functions, the maximum number of features that are selected for the best split is defined by the \textit{auto}, \textit{$log2$}, and the \textit{square root} of the number of features in the training set, the warm start parameter is set to \textit{false} and \textit{true}, the number of trees in the forest is searched in $[5, 50]$ with a gap of $5$ increasing at each step and the performance on the test set is evaluated by checking the scoring of each performance metrics. 

\textbf{SVM} classifier has \textit{radial basis function} (rbf) and \textit{linear} kernel functions with the regularization parameter searched in $[1, 1000]$ with a gap of $\times10$ increasing at each step. The kernel coefficients are determined in $[10^{-1}, 10^{-6}]$ with a decrease of $10^{-1}$ at each step and the scoring parameters for the testing phase are searched over each performance metric.

\textbf{k-NN} decides the best algorithm for computing the nearest neighbors automatically or with brute-search, BallTree, and KDTree algorithms by weighting each neighborhood \textit{uniformly} and \textit{inverse} of their distance. The number of neighbors is determined in $[5, 30]$ with a gap of $5$ increasing at each step, the metric used for computing the distances between neighbors are \textit{Manhattan} and \textit{Euclidean}, and the scoring parameters are selected as the performance metric with the highest scoring value. 

\textbf{1D-CNN} is trained by Adam optimization algorithm \cite{kingma2014adam} along with categorical cross-entropy loss function with a learning rate of $[10^{-1}, 10^{-7}]$ decreasing at each step by $10^{-1}$. The filter sizes of $[4, 8, 12, 16, 24, 32]$ and the kernel sizes of $[3, 5, 7, 9, 11, 13, 15]$ are searched to train the model with $[25, 50, 75, 100]$ epochs by setting the scoring parameter to performance metric with the highest value. 

\begin{figure}[b!]
    \centering
    \includegraphics[width=.98\linewidth]{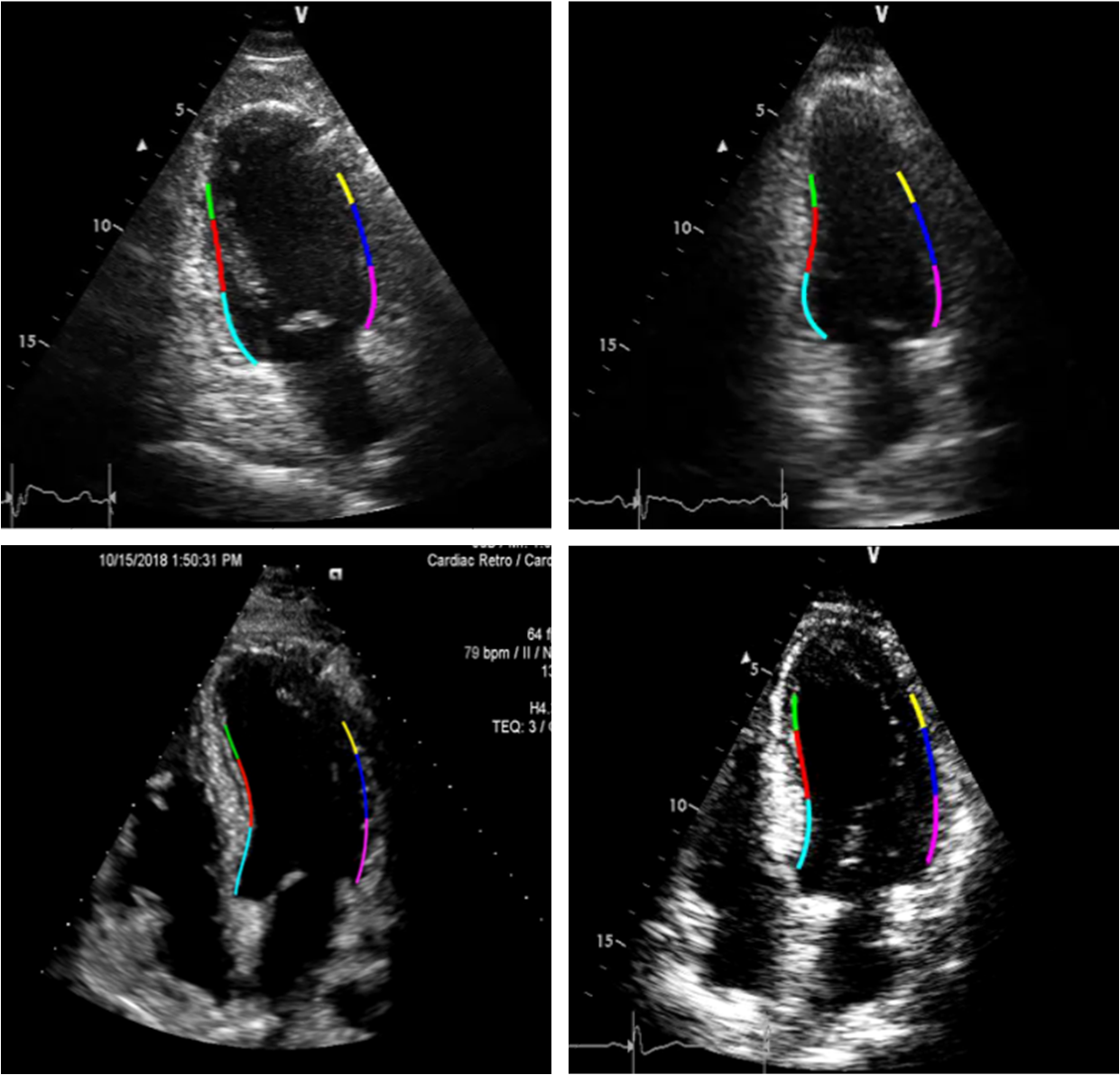}
    \caption{A4C and A2C view frames for the endocardial boundary extraction process by the APs method. The sample images at the first row are subjected to artifacts, noise, or low contrast.}
    \label{fig:APs_results}
\end{figure}

\subsection{Single-view Experimental Results}\label{sec:experimental-results-single}
In Fig. \ref{fig:APs_results}, some examples of the APs formation can be depicted. The figure reveals that the APs can successfully represent the true endocardial boundary even for low quality A4C and A2C views.

\begin{table*}[t!]
\centering
\caption{Average MI detection performance results (\%) of state-of-the-art models, threshold-based APs, and our proposed approach computed from 5-folds in A4C view echocardiography.}
\resizebox{.98\linewidth}{!}{
\begin{tabular}{cccccccc}
\hline \hline
\begin{tabular}[c]{@{}c@{}}Method\end{tabular} & Model & \multicolumn{1}{c}{Sensitivity} & \multicolumn{1}{c}{Specificity} & \multicolumn{1}{c}{Precision} & \multicolumn{1}{c}{F$1-$Score} & \multicolumn{1}{c}{F$2-$Score} & \multicolumn{1}{c}{Accuracy} \\ 
\hline \hline

\multirow{5}{*}{Kusunose \textit{et al.} \cite{kusunose2020deep}} & \cellcolor[gray]{.97}ResNet-50 & \multicolumn{1}{c}{\cellcolor[gray]{.97}$74.68$} & \multicolumn{1}{c}{\cellcolor[gray]{.97}$54.84$} & \multicolumn{1}{c}{\cellcolor[gray]{.97}$68.95$} & \multicolumn{1}{c}{\cellcolor[gray]{.97}$71.59$} & \multicolumn{1}{c}{\cellcolor[gray]{.97}$73.38$} & \multicolumn{1}{c}{\cellcolor[gray]{.97}$66.25$} \\

 & DenseNet-121 & \multicolumn{1}{c}{$73.39$} & \multicolumn{1}{c}{$63.52$} & \multicolumn{1}{c}{$73.48$} & \multicolumn{1}{c}{$72.46$} & \multicolumn{1}{c}{$72.79$} & \multicolumn{1}{c}{$69.38$}\\
 
 & \cellcolor[gray]{.97}InceptionResNetv2 & \multicolumn{1}{c}{\cellcolor[gray]{.97}$72.34$} & \multicolumn{1}{c}{\cellcolor[gray]{.97}$65.05$} & \multicolumn{1}{c}{\cellcolor[gray]{.97}$73.56$} & \multicolumn{1}{c}{\cellcolor[gray]{.97}$72.26$} & \multicolumn{1}{c}{\cellcolor[gray]{.97}$72.15$} & \multicolumn{1}{c}{\cellcolor[gray]{.97}$69.38$}  \\
 
 & \multicolumn{1}{c}{Inception-v3} & $67.95$ & $66.70$ & $73.71$ & $70.17$ & $68.72$ & $67.50$ \\
 
 & \multicolumn{1}{c}{\cellcolor[gray]{.97}Xception} & \cellcolor[gray]{.97}$85.61$ & \cellcolor[gray]{.97}$47.36$ & \cellcolor[gray]{.97}$69.26$ & \cellcolor[gray]{.97}$76.24$ & \cellcolor[gray]{.97}$81.49$ & \cellcolor[gray]{.97}$69.38$  \\ 
 \hline
 
\multirow{1}{*}{Kiranyaz \textit{et al.} \cite{9261387}} & \multicolumn{1}{c}{APs\cite{9261387}} & $86.25$ & $77.08$ & $86.25$ & $86.25$ & $86.25$ & $82.81$ \\
\hline

\multirow{5}{*}{Ours} & \cellcolor[gray]{.97}DT & \multicolumn{1}{c}{\cellcolor[gray]{.97}$85.00$} & \multicolumn{1}{c}{\cellcolor[gray]{.97}$68.00$} & \multicolumn{1}{c}{\cellcolor[gray]{.97}$80.95$} & \multicolumn{1}{c}{\cellcolor[gray]{.97}$82.93$} & \multicolumn{1}{c}{\cellcolor[gray]{.97}$84.16$} & \multicolumn{1}{c}{\cellcolor[gray]{.97}$78.46$} \\

 & RF & \multicolumn{1}{c}{$86.25$} & \multicolumn{1}{c}{$\textbf{84.00}$} & \multicolumn{1}{c}{$\textbf{89.61}$} & \multicolumn{1}{c}{$\textbf{87.90}$} & \multicolumn{1}{c}{$86.90$} & \multicolumn{1}{c}{$\textbf{85.38}$} \\
 
 & \cellcolor[gray]{.97}SVM & \multicolumn{1}{c}{\cellcolor[gray]{.97}$82.50$} & \multicolumn{1}{c}{\cellcolor[gray]{.97}$70.00$} & \multicolumn{1}{c}{\cellcolor[gray]{.97}$81.48$} & \multicolumn{1}{c}{\cellcolor[gray]{.97}$81.99$} & \multicolumn{1}{c}{\cellcolor[gray]{.97}$82.29$} & \multicolumn{1}{c}{\cellcolor[gray]{.97}$77.69$}  \\
 
 & \multicolumn{1}{c}{k-NN} & $\textbf{88.75}$ & $78.00$ & $86.59$ & $87.65$ & $\textbf{88.31}$ & $84.62$ \\
  
 & \multicolumn{1}{c}{\cellcolor[gray]{.97}1D-CNN} & \cellcolor[gray]{.97}$83.75$ & \cellcolor[gray]{.97}$78.00$ & \cellcolor[gray]{.97}$85.90$ & \cellcolor[gray]{.97}$84.81$ & \cellcolor[gray]{.97}$84.17$ & \cellcolor[gray]{.97}$81.54$  \\ 
 
 \hline
\end{tabular}}
\label{tab:experiments_SOTA}
\end{table*}

We present the performance of each classifier in single-view (A4C or A2C view) echocardiography, individually. The MI detection results are presented in Table \ref{tab:experiments_1}. In A4C view echocardiography, the prior approach with the threshold-based APs method \cite{9261387} achieves $86.25\%$ sensitivity with a specificity level of $77.08\%$. The results indicate that imposing ML into the algorithm generally outperforms the threshold-based APs method in \cite{9261387} by the classifiers utilized in this study. In the A4C view, the k-NN classifier achieves the highest sensitivity level of $88.75\%$, whereas the highest specificity of $84\%$ is obtained by the RF classifier. 

The performance of prior work in \cite{9261387} for A2C view echocardiography is $69.12\%$ sensitivity with $59.68\%$ specificity as shown in Table \ref{tab:experiments_1}. Once again, it was generally outperformed by the proposed approach with the evaluated classifiers. The SVM classifier achieved the highest sensitivity level of $76.47\%$, whereas the highest F$1-$Score is obtained by the k-NN classifier with $74.81\%$ in A2C view echocardiography.

For further investigation, we compare our proposed approach with an end-to-end solution using Deep Learning models. Accordingly, we adapted the method proposed by Kusunose \textit{et al.} \cite{kusunose2020deep} that utilizes Deep Convolutional Neural Networks (DCNNs) to detect RWMAs over circular view echocardiography recordings. For the comparison, we used A4C view echocardiography recordings of HMC-QU dataset. From state-of-the-art DCNNs, we use the models ResNet-50 \cite{he2016deep}, Inception-v3 \cite{szegedy2016rethinking}, DenseNet-121 \cite{huang2017densely}, Xception \cite{8099678}, and InceptionResNetv2 \cite{Szegedy2017} with transfer learning, where the networks are initialized with ImageNet datasetweights. We resized the echocardiography frames and selected the three frames corresponding to a one-cardiac-cycle: end-diastole to end-systole and end-systole to end-diastole to use as the input of pre-trained models with $(224\times224\times3)$ image sizes. Then, we performed a stratified $5$-fold cross-validation scheme and applied data augmentation using the Image Data Generator in Keras to the training set of each fold to avoid overfitting with an increased number of images up to $2000$. Accordingly, we compare the average MI detection performances between Kusunose \textit{et al.} \cite{kusunose2020deep} (DCNNs), Kiranyaz \textit{et al.} \cite{9261387} (threshold-based APs), and our proposed method with Machine Learning-based APs, where the performance of each model is computed over a cross-validation scheme with the same train/test sets for each fold. Consequently, Table \ref{tab:experiments_SOTA} reveals that our proposed approach outperforms the end-to-end network solution proposed by Kusunose \textit{et al.} \cite{kusunose2020deep} and achieves the highest performance for each performance metric.

\begin{table}[b!]
\centering
\caption{The confusion matrices of MI detection in multi-view echocardiography by the RF model, where the symbol $^\star$ indicates the concatenated features.}
\begin{subtable}{\linewidth}
\centering
\caption{$\text{Multi-view}^\star$}
\small
    \begin{tabular}{|c|c|c|c|}
\hline
\multicolumn{2}{|c|}{\multirow{2}{*}{\normalsize{$\text{\textbf{Multi-view}}^\star$}}} & \multicolumn{2}{c|}{\normalsize{Predicted}} \\ \cline{3-4} 
\multicolumn{2}{|c|}{} & \multicolumn{1}{c|}{non-MI} & \multicolumn{1}{c|}{MI} \\ \hline
\multirow{2}{*}{\begin{tabular}[c]{@{}c@{}}\normalsize{Ground}\\ \normalsize{Truth}\end{tabular}} & non-MI & $26$ & $16$ \\ \cline{2-4} 
 & MI & $11$ & $77$ \\ \hline
\end{tabular}
\label{CM1}
\end{subtable}

\bigskip
\noindent
\begin{subtable}{\linewidth}
\centering
\caption{Multi-view}
\small
    \begin{tabular}{|c|c|c|c|}
\hline
\multicolumn{2}{|c|}{\multirow{2}{*}{\normalsize{\textbf{Multi-view}}}} & \multicolumn{2}{c|}{\normalsize{Predicted}} \\ \cline{3-4} 
\multicolumn{2}{|c|}{} & \multicolumn{1}{c|}{non-MI} & \multicolumn{1}{c|}{MI} \\ \hline
\multirow{2}{*}{\begin{tabular}[c]{@{}c@{}}\normalsize{Ground}\\ \normalsize{Truth}\end{tabular}} & non-MI & $30$ & $12$ \\ \cline{2-4} 
 & MI & $12$ & $76$ \\ \hline
\end{tabular}
\label{CM2}
\end{subtable}
\label{tab:CM-RF}
\end{table}

\subsection{Multi-view Experimental Results}\label{sec:experimental-results-multi}
In multi-view echocardiography, we merge the single-view information by concatenating the features as $\mathbf{F} = \begin{bmatrix} \mathbf{\Phi_1}^\text{T} \text{ } \mathbf{\Phi_2}^\text{T} \end{bmatrix} \in \mathbb{R}^{1\times12}$. Alternatively, we utilize both of the single-view echocardiography results to detect MI in multi-view echocardiography by simply merging the A4C and A2C view detection results with the "\textit{OR}" operator as a straight-forward solution. Accordingly, if either of the single-view detection outcomes is MI, the multi-view detection outcome will also be MI.

\begin{figure}[b!]
    \centering
    \includegraphics[width= \linewidth]{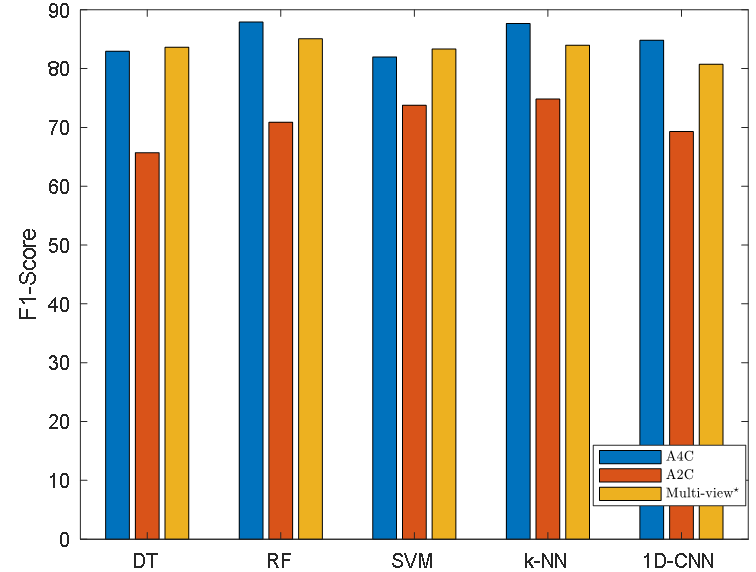}
    \caption{The F$1-$Scores of the ML classifiers are plotted for single-view (A4C and A2C) and multi-view (the proposed feature concatenation) echocardiography.}
    \label{fig:F1_plots}
\end{figure}

\begin{table*}[t!]
\centering
\caption{Average MI detection performance results (\%) computed from 5-folds in multi-view echocardiography, where the symbol $^\star$ indicates the concatenated features.}
\resizebox{.85\linewidth}{!}{
\begin{tabular}{cccccccc}
\hline \hline
\begin{tabular}[c]{@{}c@{}}Echocardiography \\ View\end{tabular} & Model & \multicolumn{1}{c}{Sensitivity} & \multicolumn{1}{c}{Specificity} & \multicolumn{1}{c}{Precision} & \multicolumn{1}{c}{F$1-$Score} & \multicolumn{1}{c}{F$2-$Score} & \multicolumn{1}{c}{Accuracy} \\ 
\hline \hline

\multirow{5}{*}{$\text{\normalsize{Multi-view}}^\star$} & \cellcolor[gray]{.97}DT & \multicolumn{1}{c}{\cellcolor[gray]{.97}$84.09$} & \multicolumn{1}{c}{\cellcolor[gray]{.97}$64.29$} & \multicolumn{1}{c}{\cellcolor[gray]{.97}$83.15$} & \multicolumn{1}{c}{\cellcolor[gray]{.97}$83.62$} & \multicolumn{1}{c}{\cellcolor[gray]{.97}$83.90$} & \multicolumn{1}{c}{\cellcolor[gray]{.97}$77.69$} \\

 & RF & \multicolumn{1}{c}{$87.50$} & \multicolumn{1}{c}{$61.90$} & \multicolumn{1}{c}{$82.80$} & \multicolumn{1}{c}{$\textbf{85.08}$} & \multicolumn{1}{c}{$86.52$} & \multicolumn{1}{c}{$\textbf{79.23}$}\\
 
 & \cellcolor[gray]{.97}SVM & \multicolumn{1}{c}{\cellcolor[gray]{.97}$\textbf{90.91}$} & \multicolumn{1}{c}{\cellcolor[gray]{.97}$42.86$} & \multicolumn{1}{c}{\cellcolor[gray]{.97}$76.92$} & \multicolumn{1}{c}{\cellcolor[gray]{.97}$83.33$} & \multicolumn{1}{c}{\cellcolor[gray]{.97}$\textbf{87.72}$} & \multicolumn{1}{c}{\cellcolor[gray]{.97}$75.38$}  \\
 
 & \multicolumn{1}{c}{k-NN} & $86.36$ & $59.52$ & $81.72$ & $83.98$ & $85.39$ & $77.69$ \\
 
 & \multicolumn{1}{c}{\cellcolor[gray]{.97}1D-CNN} & \cellcolor[gray]{.97}$76.14$ & \cellcolor[gray]{.97}$\textbf{73.81}$ & \cellcolor[gray]{.97}$\textbf{85.90}$ & \cellcolor[gray]{.97}$80.72$ & \cellcolor[gray]{.97}$77.91$ & \cellcolor[gray]{.97}$75.38$  \\ 
 \hline

\multirow{5}{*}{\normalsize{Multi-view}} & DT & $\textbf{90.91}$ & $45.24$ & $77.67$ & $83.77$ & $87.91$ & $76.15$ \\

 & \cellcolor[gray]{.97} RF & \cellcolor[gray]{.97}$86.36$ & \cellcolor[gray]{.97}$\textbf{71.43}$ & \cellcolor[gray]{.97}$\textbf{86.36}$  & \cellcolor[gray]{.97}$86.36$  & \cellcolor[gray]{.97}$86.36$  & \cellcolor[gray]{.97}$\textbf{81.54}$  \\
 
 & SVM & $88.64$ & $47.62$ & $78.00$ & $82.98$ & $86.28$ & $75.38$ \\
 
 & \multicolumn{1}{c}{\cellcolor[gray]{.97} k-NN} & \cellcolor[gray]{.97}$89.77$  & \cellcolor[gray]{.97}$64.29$  & \cellcolor[gray]{.97}$84.04$& \cellcolor[gray]{.97}$\textbf{86.81}$  & \cellcolor[gray]{.97}$\textbf{88.57}$  & \cellcolor[gray]{.97}$\textbf{81.54}$  \\
 
 & \multicolumn{1}{c}{1D-CNN} & $84.19$ & $64.29$ & $83.15$ & $83.62$ & $83.90$ & $77.69$\\
 
 \hline
\end{tabular}}
\label{tab:experiments_2}
\end{table*}

In Table \ref{tab:experiments_2}, the MI detection performances for multi-view echocardiography are presented. The results indicate that the proposed approach that concatenates the single-view features has proximate performance compared to the alternative solution in multi-view MI detection. Accordingly, the proposed multi-view approach with the SVM classifier achieves an elegant sensitivity level of $90.91\%$. On the other hand, the alternative approach with the RF classifier has the highest precision by $86.36\%$. Accordingly, the confusion matrices of the RF model that gives the highest accuracy in multi-view echocardiography are shown in Table \ref{tab:CM-RF}, where both solutions are compared. The F$1-$Score performance metrics of each ML classifier is plotted in Fig. \ref{fig:F1_plots} for single-view (A4C and A2C) and multi-view echocardiography. As it can be seen from the figure, the detection of MI is successful with the proposed approach. Due to the different myocardial segments appearing on each view, a direct comparison of the experimental results of multi-view and single-view is not viable, e.g., consider the case that MI detection from a single-view will always fail if the infarcted segment is not one of the segments visible on that view. Therefore, the results of single-view indicate the MI detection performance over the infarcted segments visible on that view only. Similarly, the results of multi-view indicate the MI detection performance over the infarcted segments visible on one of the views. Therefore, with fewer cases, the single-view performance may be higher than the one for multi-view. Hence, the reliable way to perform the diagnosis would be to use multi-view echocardiography that includes more information regarding the myocardial segment motion from both A4C and A2C views.

\subsection{Computational complexity}

The computational complexity of the proposed multi-view MI detection method is the total computational complexity that arises from each individual block depicted in Fig. \ref{fig:method}. The endocardial boundary extraction, myocardial segment displacement, and feature engineering blocks of the method are from the prior work, where their computational complexities are detailed in \cite{9261387}. However, the time elapsed for executing the algorithm is doubled since both A4C and A2C views are used in this study. On the other hand, the MI detection stage has a computational complexity that differs with respect to the utilized classifiers. Accordingly, the classifiers have the computational complexities in the prediction phase as follows: DT as $O(V)$, RF as $O(Vn_{tree})$, SVM as $O(Vn_{sv})$, and k-NN as $O(Vn_{train})$, where the length of the feature vector, number of trees, number of support vectors, and number of training samples are denoted as $V$, $n_{tree}$, $n_{sv}$, and $n_{train}$, respectively. Furthermore, the convolutional layer computations of 1D-CNN are as follows:
    \begin{equation}
        \begin{split}
        C =\sum_{l=1}^L N_{(l-1)} N_l V_{(l-1)} K_{(l-1)}^2 + \\ \sum_{l=0}^{L-1}N_{(l+1)}N_l(K_l+V_l)K_l^2 + \\
        \sum_{l=0}^{L-1}N_{(l+1)}N_lK_l(K_l+V_l)^2, 
        \label{eq:11b}
        \end{split}
    \end{equation}
\noindent where $C$ in Eq. (\ref{eq:11b}) is the multiplication operations of $L$ number of layers at each back propagation iteration, $N$ number of connections between layers, and $K$ sized filter. Thus, 1D-CNN has the time complexity $O(C)$.

Table \ref{tab:computation} shows the average time elapsed in seconds (\textit{s}) during the inference of each step of the proposed algorithm in multi-view echocardiography that is illustrated in Fig. \ref{fig:method}. Accordingly, the most time-consuming stage arises from APs, where around $60$ seconds have passed for its execution. On the other hand, the fastest block of the algorithm is MI detection with real-time execution. Overall, the proposed algorithm requires $61.0318$ seconds on average to process A4C and A2C echocardiography views with one-cardiac cycle each ($\approx30-50$ frames in total).

\begin{table}[t!]
\centering
\caption{The average time elapsed for executing the algorithm stages in multi-view echocardiography.}
\resizebox{.48\textwidth}{!}{
\begin{tabular}{|c|c|c|}
\hline
\rowcolor[gray]{0.90}Algorithm Stage & Proposed Methods & Elapsed Time (\textit{s}) \\ \hline\hline

\begin{tabular}[c]{@{}c@{}}Endocardial Boundary \\ Extraction \end{tabular} & APs & $59.0696$ \\ \hline\hline

\begin{tabular}[c]{@{}c@{}}Myocardial Segment \\ Displacement \end{tabular} & \begin{tabular}[c]{@{}c@{}}Maximum \\ Displacements \end{tabular} & $0.0712$ \\ \hline\hline

\begin{tabular}[c]{@{}c@{}}Feature \\ Engineering \end{tabular} & \begin{tabular}[c]{@{}c@{}}Scaled \\ Displacements \end{tabular} & $0.0120$ \\ \hline\hline

\multirow{5}{*}{MI Detection} & DT & $3.023\times10^{-6}$ \\
 
 & \cellcolor[gray]{0.95}RF & \cellcolor[gray]{0.95}$5.250\times10^{-5}$ \\
 & SVM & $7.592\times10^{-6}$ \\
 & \cellcolor[gray]{0.95}k-NN & \cellcolor[gray]{0.95}$3.725\times10^{-5}$\\ 
 
 & 1D-CNN & $2.665\times10^{-3}$ \\
 
 \hline
\end{tabular}}
\label{tab:computation}
\end{table}

\section{Conclusions}\label{sec:conclusion}
The early detection of MI is a crucial task to prevent further tissue damages or even death. In this study, we propose to detect MI over multi-view echocardiography by merging the information extracted from A4C and A2C views. Contrary to the recent studies proposed for single-view, this is the first study that accomplishes a multi-view MI detection for a reliable and robust diagnosis. Moreover, this study shows that the threshold-based APs method in \cite{9261387} can significantly be improved by using an ML-based approach even for single-view MI detection. The experimental results show that the detection performance has increased with the proposed approach in single-view echocardiography by $2.50\%$ and $7.35\%$ for the sensitivity metric in A4C and A2C views, respectively. Furthermore, in multi-view echocardiography, the proposed approach has achieved a sensitivity level of $90.91\%$ and an F$2-$Score of $87.72\%$.

The proposed method can be clinically used as an assistive tool to help cardiologists and technicians to prevent subjective and operator-dependent assessments by accurately measuring the LV myocardial displacements and plotting the color-coded myocardial segments. Finally, another major contribution of this study is the formation of the multi-view HMC-QU dataset that is publicly shared with the research community. We plan to extend our approach to other views in order to detect MI due to the blockage of \textit{any} coronary artery. With this accomplishment, we will be able to identify the blocked arteries and also predict the location of the blockage simultaneously for the benefit of localizing the revascularization targets.

\bibliographystyle{IEEEtran}
\bibliography{IEEEtran}
\balance

\end{document}